\documentclass[9pt,twocolumn,twoside,showpacs,superscriptaddress]{revtex4-1}
\usepackage{
  color,
  hyperref,
  graphicx,
  amsfonts,
  amsmath,
  float
}
\usepackage[utf8]{inputenc}

\definecolor{creferee1}{rgb}{0,0,0}
\definecolor{creferee2}{rgb}{0,0,0}

\newcommand{\LCPMR}{Sorbonne Universit\'e, UPMC Univ. Paris 6, CNRS-UMR 7614, Laboratoire de Chimie Physique-Mati\`ere et Rayonnement, 4 place Jussieu, 75231 Paris Cedex 05, France}
\newcommand{\TCPCI}{Theoretische Chemie, Physikalisch-Chemisches Institut, Universit\"at Heidelberg, Im Neuenheimer Feld 229, D-69120 Heidelberg, Germany}
\newcommand{\DPA}{Department of physics and astronomy, Aarhus University, 8000 Aarhus C, Denmark}
\newcommand{\LIDYL}{LIDYL, CEA, CNRS, Universit\'e Paris-Saclay, CEA-Saclay, 91191 Gif sur Yvette, France}
\newcommand{\MPI}{Max-Planck-Institut f\"ur Kernphysik, Saupfercheckweg 1, 69117, Heidelberg, Germany}

\begin{document}

\title{Laser-induced distortion of structural interferences in high harmonic generation}

\begin{abstract}
We study theoretically the two-center interferences occurring in high harmonic generation from diatomic molecules.
By solving the time-dependent Schr\"odinger equation, either numerically or with the molecular strong-field approximation, we show that the electron dynamics results in a strong spectral smoothing of the interference, depending on the value and sign of the driving laser field at recombination. Strikingly, the associated phase-jumps are in opposite directions for the short and long trajectories. In these conditions, the harmonic emission is not simply proportional to the recombination dipole any more. This has important consequences in high harmonic spectroscopy, e.g., for accessing the molecular-frame recombination dipole in amplitude and phase.
\end{abstract}


\author{Fran\c{c}ois Risoud}
\affiliation{\LCPMR}

\author{Camille L\'ev\^eque}
\affiliation{\LCPMR}
\affiliation{\TCPCI}
\affiliation{\DPA}

\author{Marie Labeye}
\affiliation{\LCPMR}

\author{J\'er\'emie Caillat}
\affiliation{\LCPMR}

\author{Alfred Maquet}
\affiliation{\LCPMR}

\author{Pascal Sali\`eres}
\affiliation{\LIDYL}

\author{Richard Ta\"ieb}
\email{corresponding author: richard.taieb@upmc.fr}
\affiliation{\LCPMR}

\author{Tahir Shaaran}
\affiliation{\LIDYL}
\affiliation{\MPI}

\date{Compiled \today}

\maketitle


High Harmonic Spectroscopy (HHS) is a powerful technique in which the process of High Harmonic Generation (HHG) is used to probe the structure and dynamics of the generating medium with Angstr\"om and attosecond resolution. In HHG, an Electron Wave-Packet (EWP) is prepared by tunnel ionization of the atomic/molecular target gas, it is then accelerated by the strong laser field and finally driven back to the core \cite{corkum_plasma_1993,schafer_above_1993}. In the recombination process, a burst of extreme ultraviolet photons is emitted, encoding a wealth of information on the target. First, the recombination dipole moment imprints the structure of the molecular orbital involved in the emission, in the form of structural interferences \cite{lein_role_2002} equivalent to Cohen-Fano interferences in the photo-ionization dipole \cite{cohen_interference_1966}. This allows performing tomographic reconstruction of the orbital \cite{itatani_tomographic_2004,haessler_attosecond_2010,vozzi_generalized_2011}. Second, any dynamics occurring in the core during the EWP continuum excursion strongly affects the recombination. For instance, tunnel ionization from different valence orbitals results in interfering channels in the recombination \cite{McFarland_high_2008,smirnova_high_2009}.

However, it is crucial to disentangle the different effects in order to access this rich information. This was generally performed by recording the dependence on the laser parameters (intensity, wavelength), since it was widely accepted that structural interferences do not depend on them \cite{worner_controlling_2010}. This appears clearly in the Quantitative ReScattering theory (QRS) \cite{le_quantitative_2009} where the harmonic dipole is expressed as the product of the returning EWP, containing all the laser parameters, by the target-specific recombination dipole, only dependent on the EWP energy and recollision direction. Such a factorization has been very successful in predicting and explaining resonant features in the harmonic intensity, e.g., in \cite{shiner_probing_2011}. It was further confirmed by analytic calculations in the case of a model potential \cite{frolov_analytic_2011}.

In this letter, we study the signatures of the electron {\it dynamics} on the {\it structural} interference occurring in the harmonic emission from diatomic molecules. Based on Time-dependent Schr\"odinger Equation (TDSE) calculations resolving the two shortest EWP trajectories, we show that the value of the laser field at recombination plays a crucial role on the shape of the two-center destructive interference. In most cases, the intensity minimum and the associated phase-jump are spread over a large spectral range resulting in a very smooth behavior, and the phase-jumps for the short and long EWP trajectories are in opposite directions. Furthermore, we identify very peculiar conditions for which this phase-jump is very sharp.

Using the Strong-Field Approximation (SFA) approach \cite{lewenstein_theory_1994} {\it modified} for molecules \cite{chirila_strong-field_2006, faria_high-order_2007}, we explain these features by the modification of the saddle-point trajectories induced by the fast phase variation of the recombination dipole. We finally propose a method for evidencing the opposite and smoothed phase variations of the short/long contributions using quantum path interferences.
Our findings thus question the {\em strict} separation between continuum dynamics and recombination advocated, e.g., in QRS, with strong implications for HHS.

{\color{creferee2}
Our study is performed on a one dimensional (1D) system that contains the essential physics of two-center interference in HHG, while avoiding EWP spreading that was previously invoked to explain phase smoothing effects \cite{chirila_explanation_2009,van_der_zwan_two-center_2010}. Our model system represents a diatomic molecule with nuclei fixed, under the single-active electron approximation.

First, we solved the TDSE to simulate the ``exact'' electron dynamics under the influence of a low-frequency strong laser field. The electron-nuclei interaction is represented by:
}
\begin{equation}\label{eq_potential}
    V(x)=\frac{1}{2}\left[-\frac{1}{\sqrt{\left(x+\frac{R}{2}\right)^2+a^2}}-\frac{1}{\sqrt{\left(x-\frac{R}{2}\right)^2+a^2}}\right],
\end{equation}
where $R$ is the internuclear distance. This is valid for molecules with ``heavy'' nuclei that exhibit a strong two-center character, such as CO$_2$ \cite{boutu_coherent_2008,vozzi_controlling_2005,vozzi_generalized_2011}. In order to study the $R$-dependence of the structural interferences, the regularization parameter $a$ is adapted to maintain the same ground state energy ($-I_p=-0.567$ a.u.$=-15.43$ eV) at each considered value of $R$. {\color{creferee2} Varying $R$ in 1D is essentially equivalent to changing the alignment angle $\theta$ of the molecule with respect to the laser polarization that results, for the two-center interference, in an effective internuclear distance $R\cos\theta$.} {\color{creferee1} The electric field $E(t)$, of amplitude $E_L$, has a sine-square envelop lasting two optical cycles with frequency $\omega_L$ corresponding to a 800-nm Ti:sapphire laser. The carrier envelop phase is set equal to zero in order to obtain the contributions of only one set of short and long trajectories in the HHG spectrum.}
The $t\rightarrow\omega$ Fourier transform of the acceleration \cite{burnett_calculation_1992} provides the harmonic field.
We calibrate the latter with the one calculated for a reference ``atom'' ($R=0$ in Eq. \ref{eq_potential}) with same ionization potential. {\color{creferee2} This standard procedure in HHS allows removing both the plateau-cutoff shape in the spectrum and the group delay dispersion associated with the attochirp \cite{mairesse_attosecond_2003}, in order to evidence the two-center signatures we are interested in}. {\color{creferee1} We numerically discriminate the short and long trajectory contributions using an absorber at a distance $E_L/\omega^2$ \cite{risoud_phd_2016}, taking advantage of the fact that the short trajectories never go above this limit while the long ones always cross it \cite{corkum_plasma_1993}.}

\begin{figure}[h]
\begin{center}
  \includegraphics[]{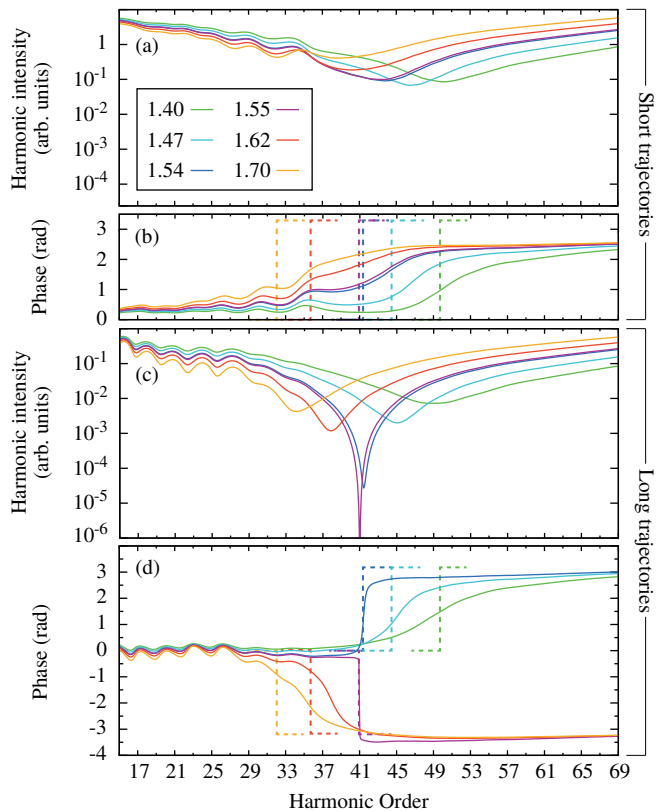}
  \caption{(Color online) TDSE computations of the short and long trajectory contributions to HHG for a two-cycle laser pulse of $2.55\times10^{14}$ W.cm$^{-2}$ peak intensity for model molecules of different internuclear distances indicated in a.u. in the legend. We report the harmonic intensity (a) [(c)] and phase (b) [(d)] calibrated by the atomic reference for the short [long] trajectories {\color{creferee1}(solid lines), and the phase of the transition dipole matrix element computed numerically for the different internuclear distances [dashed lines in (b) and (d)].}}
  \label{fig_TDSE_ampl_phase_R}
\end{center}
\end{figure}

Figure \ref{fig_TDSE_ampl_phase_R} reports the calibrated harmonic intensity and phase for several internuclear distances between 1.4 and 1.7 a.u., and a laser peak intensity of $2.55\times10^{14}$ W.cm$^{-2}$.
As expected \cite{lein_role_2002}, the possibility for the electron to recombine on either center results in a destructive two-center interference that appears as a minimum in the spectrum accompanied by a $\approx \pi$-rad phase-jump. Its position moves to high harmonic orders with decreasing $R$, since it requires a shorter EWP de Broglie wavelength $\lambda_{dB}$ ($R\approx \lambda_{dB}/2$).
However, we observe in Fig.~\ref{fig_TDSE_ampl_phase_R} many remarkable unexpected features: i) for most $R$ values, the intensity minimum and associated  phase-jump are spread over a large spectral range; {\color{creferee2} ii) the phase-jumps for the short and long trajectories are in opposite directions for high values of $R$;} iii) the shape of the interference strongly depends on $R$. In particular for the long trajectory, at a critical internuclear distance $R_c\approx 1.55$ a.u., the interference becomes extremely destructive at harmonic 41 (H41) with a very steep $\pi$ phase-jump. For smaller $R$ values, the interference becomes smooth again, but with a change of sign of the phase-jump that now resembles that of the short trajectory. Indeed, the interference is now positioned in the cutoff region ($>$H41) where the two trajectories coalesce.

Within the QRS, the calibrated emission should be directly proportional to the molecular recombination dipole matrix element. We computed it with the numerically \textit{exact} scattering waves of the potential $V(x)$ and found an excellent agreement \textit{only} at the critical distance $R_c$ and for the long trajectory.

In order to shed light on the above features, we performed a time-frequency analysis of the harmonic dipole using Gabor transforms as in \cite{antoine_time_1995}. We retrieved the emission times of the harmonics for which the destructive interferences occur and found that, at the particular internuclear distance $R_c$, it corresponds to an almost zero instantaneous electric field for the long trajectory. Furthermore, the greater (lower) values of $R$ lead to recollision with a negative (positive) electric field for the long trajectories while it is always positive for the short trajectories. Thus the sign and shape of the phase-jump seem to be strongly correlated to the instantaneous value of the electric field when the harmonics are emitted.

\begin{figure}[h]
\begin{center}
  \includegraphics[]{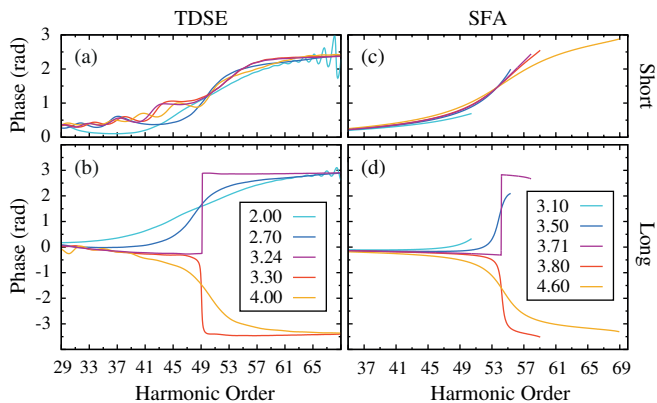}
  \caption{(Color online) TDSE [Molecular SFA] computations of the harmonic phase of our model molecule ($R=1.4$ a.u.) relative to an atomic reference for the short (a) [(c)] and long trajectories (b) [(d)] at various intensities of a two-cycle laser pulse, indicated in units of $10^{14}$ W.cm$^{-2}$ in the legend.}
  \label{fig_phase_S_L_I}
\end{center}
\end{figure}

To further examine the influence of the electric field, we varied the peak laser intensity $I_L$ from 2 to 4$\times10^{14}$ W.cm$^{-2}$ at fixed $R=1.4$ a.u. and show the results in Figs. \ref{fig_phase_S_L_I} (a,b). While the position of the destructive interference is relatively independent on $I_L$ (i.e., around H49), the phase-jump behaves similarly as when varying $R$ at fixed $I_L$. We found a critical intensity $I_c=3.24\times10^{14}$ W.cm$^{-2}$ for which we observe the inversion of direction with a very sharp $\pi$-jump. Here again, this value directly corresponds to an instantaneous electric field $\approx 0$ when H49 is emitted.

We carried out the same study within the molecular SFA framework, derived following \cite{faria_high-order_2007}, the results of which are reported in Fig.~\ref{fig_phase_S_L_I}(c,d). The same conclusions concerning the phase-jump can be drawn, with minor differences on the position of the destructive interference and the laser intensity range. These differences are due to the way the continuum is described in SFA [plane-wave (PW) approximation], and also to the position of the cutoff beyond which harmonics cannot be computed within SFA. {\color{creferee2} These SFA computations reproduce remarkably well the smoothing observed in the TDSE simulations. This evidences that the main origin of the smoothing is not related to Coulomb scattering effects \cite{ciappina_influence_2007}.}

A major advantage of the SFA lies in its semi-analytical form. It allows for a semi-classical analysis by searching the saddle-points of the action that determine the main quantum orbits/trajectories contributing to HHG \cite{lewenstein_theory_1994,salieres_feynmans_2001}.
Thus, we performed a Taylor expansion of the stationary solutions for the molecular SFA around the ones for the reference atom to get physical insight into the electron dynamics. A detailed derivation will be presented in \cite{risoud_phd_2016}, while the main outcomes are presented below. Following \cite{chirila_strong-field_2006,faria_high-order_2007}, we approximate the {\it symmetric} electronic ground state of a homonuclear diatomic molecule by the sum of two symmetric {\it atomic} orbitals centered on each nucleus. The PW-recombination and -ionization dipole matrix elements become proportional to $\cos(p R/2)$, with $p=\sqrt{2(\Omega-I_p)}$ the continuum electron momentum associated with each harmonic frequency $\Omega$.
By expressing $\cos(p R/2)$ in terms of $\exp[\pm i p R/2]$, the action reads:
\begin{eqnarray}
  S_{\Omega}(p,t,t')&=&\int_{t'}^t d\tau\left(\frac{[p+A(\tau)]^2}{2}+I_p\right)+\Omega t\nonumber\\
  +\big\{(-1)^j[p & + & A(t')]-(-1)^k[p+A(t)]\big\}\frac{R}{2}\,,
 \end{eqnarray}
where $j,k\in\{1,2\}$ label the atomic centers and $A$ is the potential vector associated with the electric field. The first two terms are identical to the atomic SFA while the last two are induced by the fast phase variations of the molecular ionization and recombination dipoles. The saddle-point equations associated with this modified action \cite{chirila_strong-field_2006, faria_high-order_2007} allow to determine the stationary solutions for the momentum $p_{\mathrm{stat}}$, the ionization and recombination times $t'_{\mathrm{stat}}$ and $t_{\mathrm{stat}}$ respectively. Compared to their atomic equivalents, we found that $p_{\mathrm{stat}}$ is unchanged at first order in the molecular case, while the times $t'_{\mathrm{stat}}$ and $t_{\mathrm{stat}}$ differ by:
\begin{equation}
 \Delta t_{\mathrm{stat}}^{\prime}=i\frac{(-1)^j R/2}{\sqrt{2I_p}}\,\,\,\mathrm{and}\,\,\, \Delta t_{\mathrm{stat}}=\frac{(-1)^k R/2}{\sqrt{2(\Omega-I_p)}},
\end{equation}
respectively. The physical meaning of these quantities is rather straightforward. They correspond to an increase (decrease) of either the barrier to tunnel through or the path in the continuum, for the electron, whether the nucleus is located at $-R/2$ ($+R/2$) for $j$ or $k$ equal $1$ or $2$, respectively. As the former is classically forbidden, the change in ionization time is purely imaginary, while the latter, classically allowed, is purely real.

The saddle-point harmonic dipole can then be factorized in the spirit of the QRS but with a modified recombination dipole, which reads to first order:
 \begin{eqnarray}\label{eq_drec_tilde}
  \tilde{d}_{\mathrm{rec}}(p,t)&=&2i\,\mathcal{R}(p)\cos\left(p\frac{R}{2}\right) \nonumber \\
  &+&2\left[E(t)|\Delta t_{\mathrm{stat}}|\frac{\partial \mathcal{R}(p)}{\partial p}+\zeta\right]\sin\left(p\frac{R}{2}\right),
\end{eqnarray}
where $\zeta$ is a constant and $\mathcal{R}(p)$ is  proportional to the overlap of the atomic orbitals with the PWs representing the continuum \cite{risoud_phd_2016}. One clearly sees that in addition to the ``bare'' recombination dipole (first term in Eq. (\ref{eq_drec_tilde}), here purely imaginary), $\tilde{d}_{\mathrm{rec}}$ contains a real valued part varying like $\sin(pR/2)$ and depending on the electric field at recombination time $t$. {\color{creferee1} The constant $\zeta$, small compared to the other term in the bracket, reflects an additional phase which originates from the rigorous handling of the prefactors arising from the saddle-point approximation within SFA \cite{risoud_phd_2016}. It explains why the discontinuous $\pi$-jump is observed at an almost but not exactly zero electric field at recombination time. }

Therefore, (i) $\tilde{d}_{\mathrm{rec}}$  is not restricted to the imaginary axis across the HHG spectra and travels in the complex plane, (ii) it does not vanish when the two-center interference term, proportional to $\cos(pR/2)$, is zero, except when the bracket in Eq. (\ref{eq_drec_tilde}) is also zero, (iii) it has a real part which is almost proportional to $E(t)$ ($\zeta$ is significantly small with respect to the other term in the brackets \cite{risoud_phd_2016}) and which value and sign drives the direction and the smoothness of the phase-jump. Thus, $\tilde{d}_{\mathrm{rec}}=0$ happens {\it only} when $E(t)\approx 0$ at emission time  of H49 of the long trajectory, confirming our findings from both the TDSE and the full molecular SFA simulations.

One should note that, if we neglect $\zeta$  in Eq. (\ref{eq_drec_tilde}), we recover an expression for $\tilde{d}_{\mathrm{rec}}$ close to the one found by including the \textit{ad hoc} dressing of the molecular ground state by the electric field, in the light of Ref. \cite{spiewanowski_field-induced_2014}. Indeed, in the presence of a laser field, we may approximate the dressed fundamental wave-function as a linear combination:
\begin{equation}
  \Phi_0(x,t) \simeq \psi_{\sigma_g}(x)+c(t)\,\psi_{\sigma_u}(x)\,,
\end{equation}
of a symmetric state $\sigma_g$ (field-free molecular ground-state), sum of two symmetric atomic orbitals centered at $\pm R/2$, and an antisymmetric state $\sigma_u$ (field-free first excited state), difference of the same two atomic orbitals \cite{odzak_interference_2009}. This decomposition is fully valid for large internuclear distances $R$, and a good approximation at shorter distances \cite{risoud_phd_2016}.

Within the time-dependent perturbation theory, $c(t)$ depends linearly on the instantaneous electric field $E(t)$, i.e., $c(t)=\chi E(t)$, and we checked, by computing the exact dressed ground-state \cite{maier_spherical-box_1980}, that this is still valid for fields strength $\approx 10^{14}$ W.cm$^{-2}$. Thus, the modified recombination dipole associated with $ \Phi_0$ within the PW approximation, becomes:
\begin{equation}
d_{\mathrm{rec}}(p,t) \propto i\cos\left(p\frac{R}{2}\right) -\chi E(t)\sin\left(p\frac{R}{2}\right),
\end{equation}
which resembles the expression given in Eq. (\ref{eq_drec_tilde}). Thus, the signature of the electron dynamics in $\tilde{d}_{\mathrm{rec}}$ was attributed to the manifestation of the dressing of the electronic ground-state.

In practice, the influence of the electron dynamics on the structural interference could be observed by following the phase difference between the short/long trajectory contributions when varying the laser intensity, e.g., by recording the Quantum Path Interferences (QPI) in the total harmonic dipole \cite{zair_quantum_2008,zair_molecular_2013}. 

In summary, we have shown that the electron dynamics in the HHG process is modified by the presence of two-center interferences. The value and sign of the driving laser field at recombination become critical in spreading the intensity minimum and associated phase-jump over a large spectral range and inducing opposite phase-jumps for the short and long trajectories. Except in specific situations, our study demonstrates that the harmonic emission is \textit{not} proportional to the recombination dipole moment. This raises important questions since such factorization is central in HHS, e.g., in the QRS theory. Care should thus be taken when comparing the recombination dipole extracted from HHG to the photoionization dipole, or using it for molecular imaging and quantum tomography. Our conclusions could be generalized to other types of resonances occurring in the recombination dipole, like shape or autoionization resonances. If narrow enough, the corresponding fast phase variations should modify the electron dynamics and result in a smoother behavior in the harmonic emission. Further studies are needed to uncover all the implications of the reported findings.

\section*{Funding Information}
We acknowledge financial support from the LABEX Plas@Par project, the program ``Investissements d'avenir'' under the reference ANR-11-IDEX-0004-02, the program ANR-15-CE30-0001-01-CIMBAAD and the European Network ITN-ATTOFEL.

\section*{Acknowledgments}
We thank T. Auguste and T. Ruchon for enlightening discussions and early simulations. 

\bibliography{biblio}
\clearpage

\end{document}